\documentclass[12pt,preprint]{aastex}

\shorttitle{Whims of an Accreting Young Brown Dwarf}
\shortauthors{Scholz, Jayawardhana, \& Brandeker}

\begin{document}


\title{Whims of an Accreting Young Brown Dwarf:\\
Exploring Emission Line Variability of 2MASSW J1207334-393254}

\author{Alexander Scholz\footnote{aleks@astro.utoronto.ca}, 
Ray Jayawardhana\footnote{rayjay@astro.utoronto.ca}, 
\& Alexis Brandeker\footnote{brandeker@astro.utoronto.ca}}
\affil{Department of Astronomy \& Astrophysics, University of Toronto,
    60 St. George Street, Toronto, Ontario M5S3H8, Canada}

\begin{abstract}
We report the first comprehensive study of emission line variability 
in an accreting young brown dwarf. We have collected 14 high-resolution 
optical spectra of 2MASSW J1207334-393254 (M8), a likely member of the 
nearby 8-million-year-old TW Hydrae association with a recently 
identified planetary mass companion, in three observing runs between 
2005 January-March on Magellan Clay telescope. 
These spectra show a variety of emission lines that are commonly seen 
in classical T~Tauri stars. H$\alpha$ line in particular shows dramatic 
changes in shape and intensity in our dataset, both on timescales of 
several weeks and several hours. In spectra from late-January, the line 
is relatively weak and only slightly asymmetric. Spectra from mid- and 
late-March show intense, broad (10\% width $\sim$280\,$\mathrm{km\,s^{-1}}$) 
and asymmetric H$\alpha$ emission, indicative of on-going disk accretion. 
Based on empirical diagnostics, we estimate that the accretion rate could
have changed by a factor of 5-10 over $\sim$6 weeks in this
brown dwarf, which may be in the final stages of accreting from its
disk. March spectra also reveal significant `quasi-periodic' changes
in the H$\alpha$ line profile over the course of a night, from
clearly double-peaked to nearly symmetric. 
These nightly profile changes, roughly consistent with
the brown dwarf's rotation period, could be the result of a 
redshifted absorption feature coming into and out of our line of
sight; when the profile is double-peaked we may be looking into an
accretion column, flowing from the inner disk edge on to the central
object, indicating that the accretion is probably channelled along the 
magnetic field lines. Our findings provide strong support for  
the magnetospheric accretion scenario, and thus for the existence 
of large-scale magnetic fields, in the sub-stellar regime. 
We also present the first high-resolution optical spectrum
of SSSPM J1102-3431 (M8.5), which has recently been identified as
another likely sub-stellar member of the TW Hydrae association. Its
emission lines are relatively narrow and fairly symmetric, suggesting
that it is accreting only very weakly, if at all.
\end{abstract}


\section{Introduction}

Over the past few years, it has become increasingly clear that
young brown dwarfs undergo a T~Tauri phase, remarkably similar
to that of their stellar counterparts. Dusty disks appear to be
common around young brown dwarfs, as indicated by near- and
mid-infrared excess \citep[e.g.,][]{ntc02,jas03,mjn04}. Spectroscopic 
signatures of disk accretion, such as broad, asymmetric H$\alpha$ 
emission lines, are now seen in objects down to nearly the 
Deuterium-burning limit \citep[e.g.,][]{jmb02,jmb03,ntm04,mjb05}. 
Irregular large-amplitude photometric variations, observed in some 
young brown dwarfs, may be related to accretion as well \citep{se04}. 
A number of objects near and below the sub-stellar boundary also
show forbidden emission lines, usually associated with jets and
winds in T~Tauri and Class~I sources \citep[e.g.,][]{fc01,bmj04,ntm04,bj04}.

\citet{g02} identified the brown dwarf 2MASSW J1207334-393254  (hereafter
2M1207) as a likely member of the nearby $\sim$8-Myr-old TW Hydrae
association, and reported strong H$\alpha$ emission in a low-resolution
spectrum, possibly indicative of disk accretion. \citet{jas03}
did not find significant L'-band (3.8$\mu$m) excess from it, but \citet{spa04}
detected mid-infrared excess, a situation analogous to two other
stellar members of TW Hydrae \citep{jhf99}. Mohanty, Jayawardhana 
\& Barrado y Navascu\'es (2003; hereafter MJB03) 
obtained high-resolution spectra of 2M1207, which showed broad asymmetric 
H$\alpha$ as well as other strong Balmer 
and He I lines, bolstering the case for on-going accretion. These findings 
drew widespread attention to 2M1207 as a brown dwarf with a long-lived 
accretion disk that could serve as a useful benchmark for studies of disk 
evolution in the sub-stellar regime. Recently, \citet{cld04,cld05} have 
found a likely planetary mass companion to 2M1207; the pair may have formed 
as a relatively wide brown dwarf binary.

\citet{gb04} noted significant differences in the equivalent width (EW) 
of H$\alpha$ between their own 2001 measurements and those of Kirkpatrick et
al. (unpublished) and MJB03. These hints of variations, together with 
the broad interest in 2M1207, inspired us to investigate possible accretion 
variability in this system. Here we report dramatic changes in H$\alpha$ 
emission seen in a series of high-resolution spectra taken between 2005 
January-March. We also present the first high-resolution spectrum of 
SSSPM J1102-3431 (hereafter SSSPM1102), another likely sub-stellar member 
of the TW Hydrae association, recently identified by \citet{smz05}.

\section{Observations and spectral analysis}
\label{obs}

We obtained spectra of 2M1207 in three observing runs with the 
echelle spectrograph MIKE at the Magellan Clay 6.5-m telescope on
Las Campanas. In the first run in January 2005 and the third run in March 2005, 
we used no binning and 0\farcs35 slit, which gives a resolution of $R\sim 60000$. 
Integration times are between 900 and 1800\,sec. The second run, also in March 
2005, was carried out with $2\times2$ binning and 1\farcs0 slit, providing $R\sim 20000$.
The exposure times in this run are $3\times 600$\,sec. Our secondary target 
SSSPM1102 was observed on 2005 March 16 with $3\times1200$\,sec exposure
time. Summary observing log for 2M1207 is given in Table \ref{log}. 
MIKE is a double echelle spectrograph, consisting of blue and red arms, 
providing coverage from 3400 to 5000\,\AA\ and 4900 to 9300\,\AA, respectively, 
in our configuration. All spectra were reduced using standard routines within IRAF. 
Consecutive exposures were co-added to improve the signal-to-noise ratio.

\begin{figure}
\begin{center}
\includegraphics[width=3.1cm,height=3.5cm]{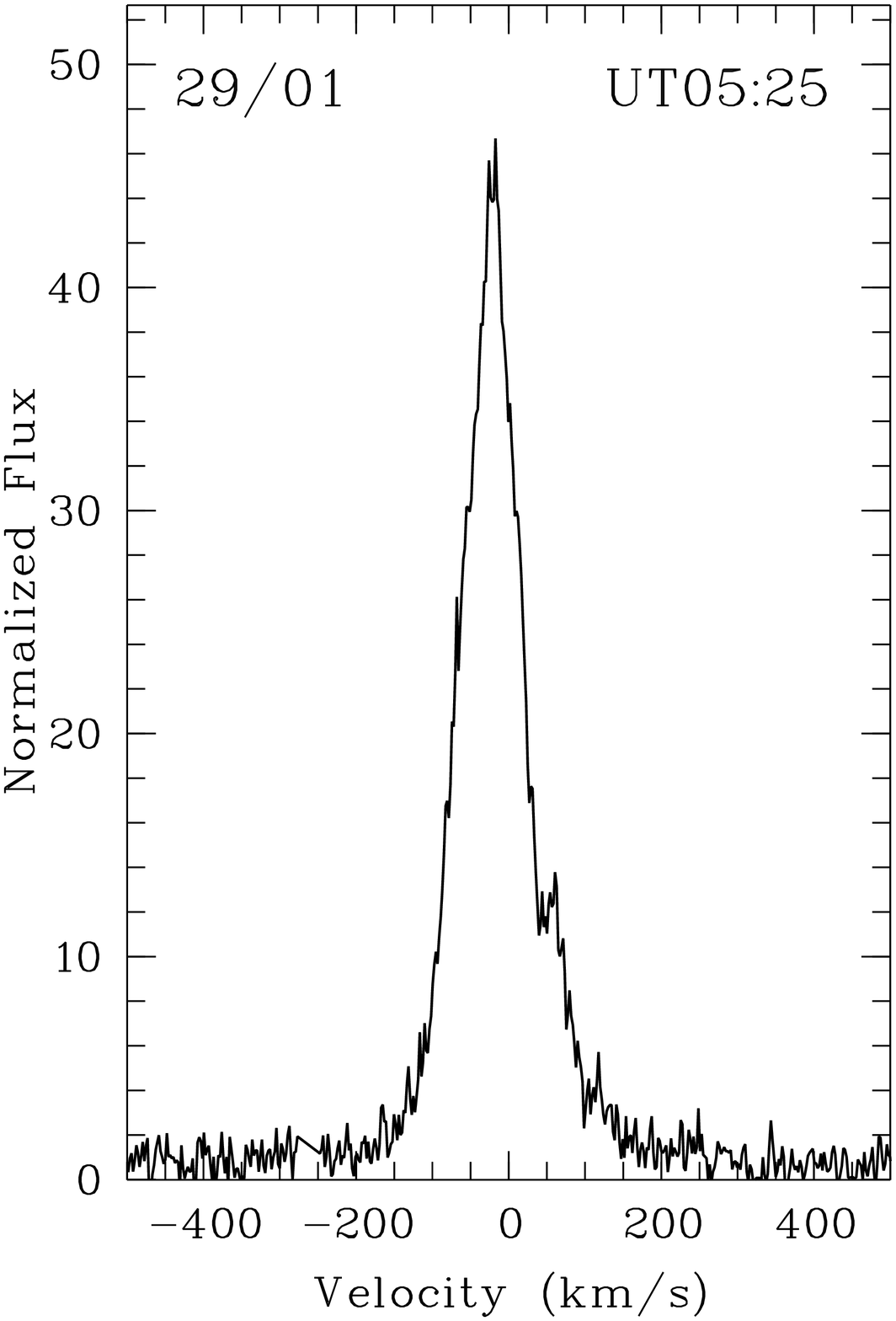}
\includegraphics[width=3.1cm,height=3.5cm]{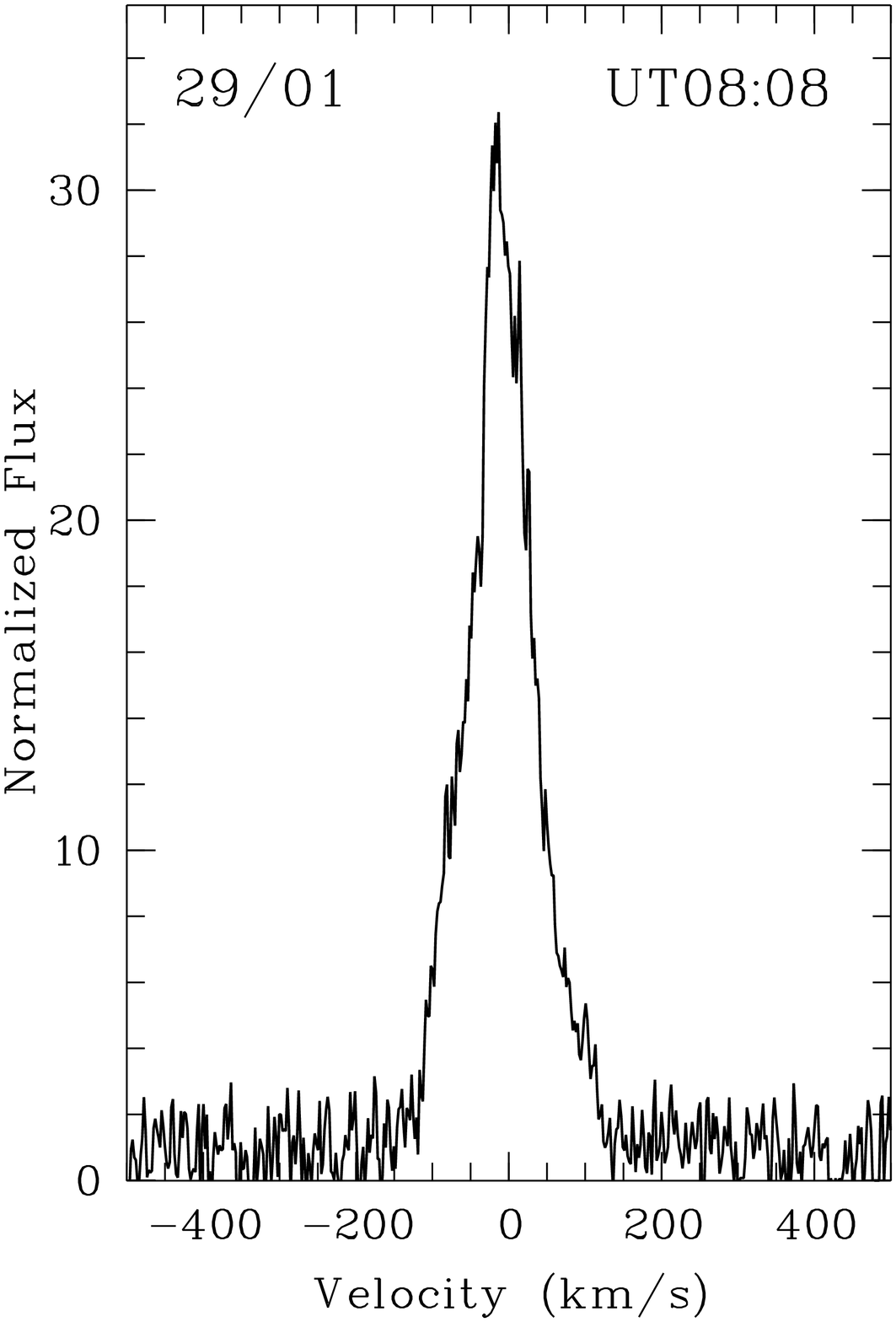}
\includegraphics[width=3.1cm,height=3.5cm]{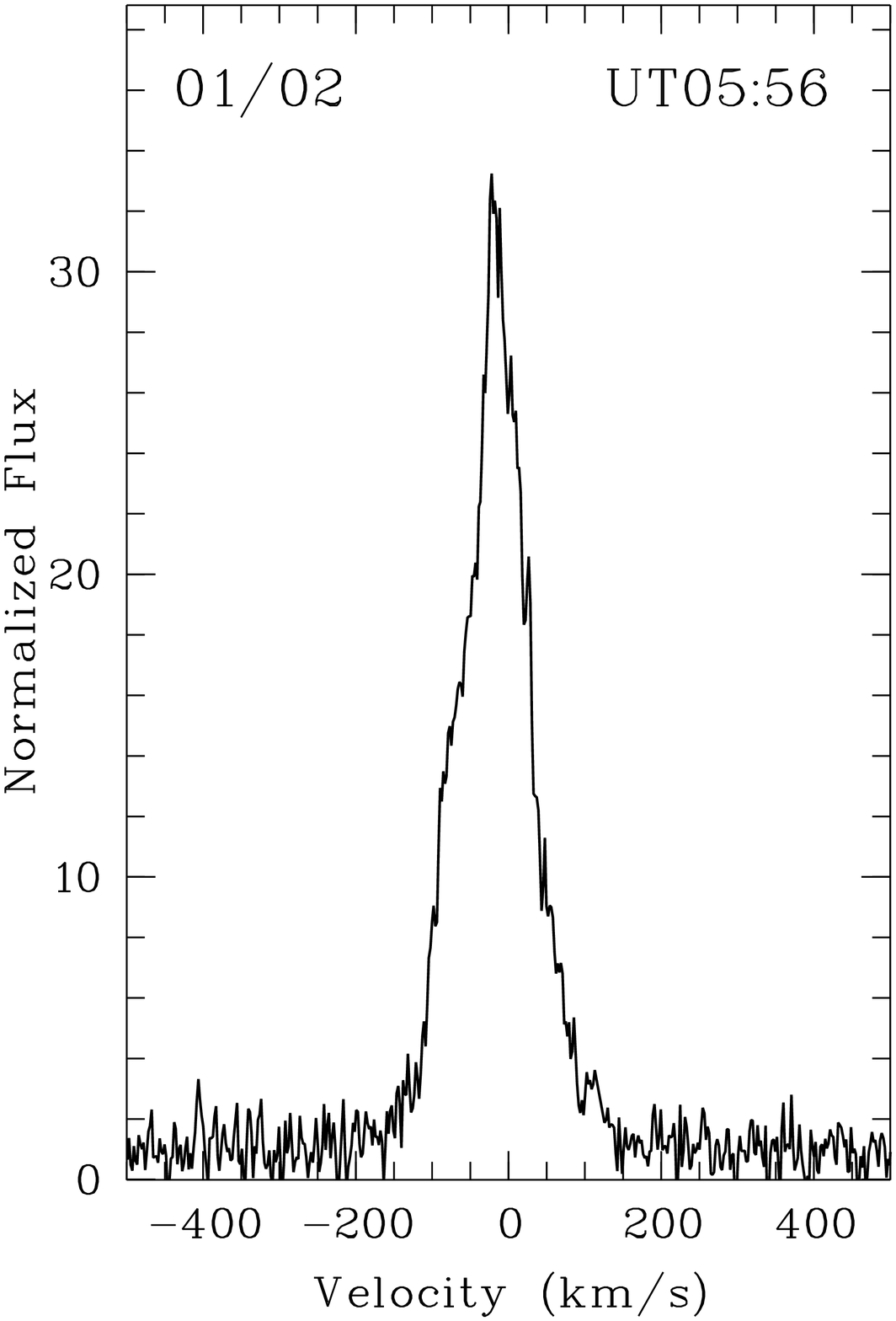}
\includegraphics[width=3.1cm,height=3.5cm]{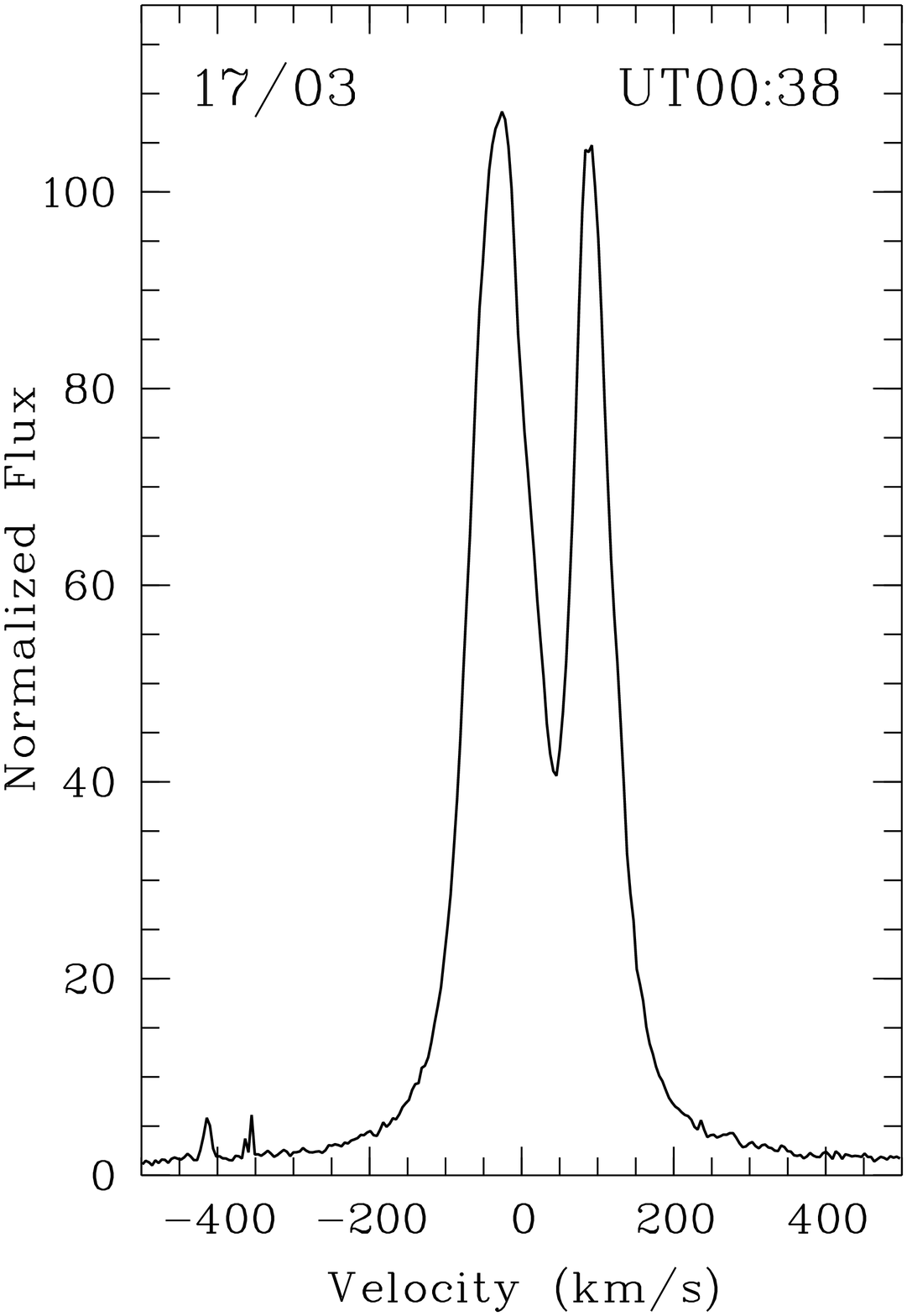}
\includegraphics[width=3.1cm,height=3.5cm]{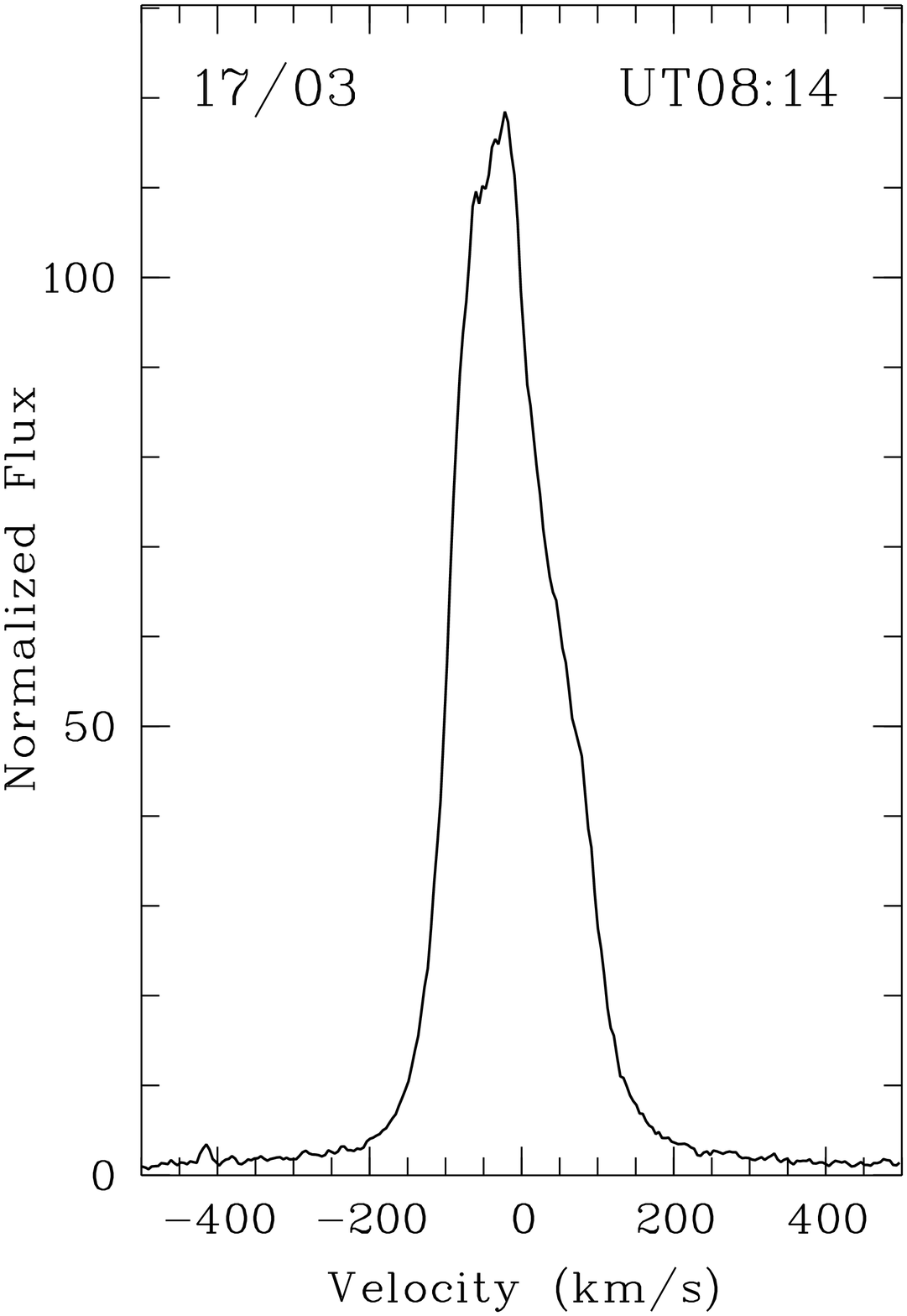}\\
\includegraphics[width=3.1cm,height=3.5cm]{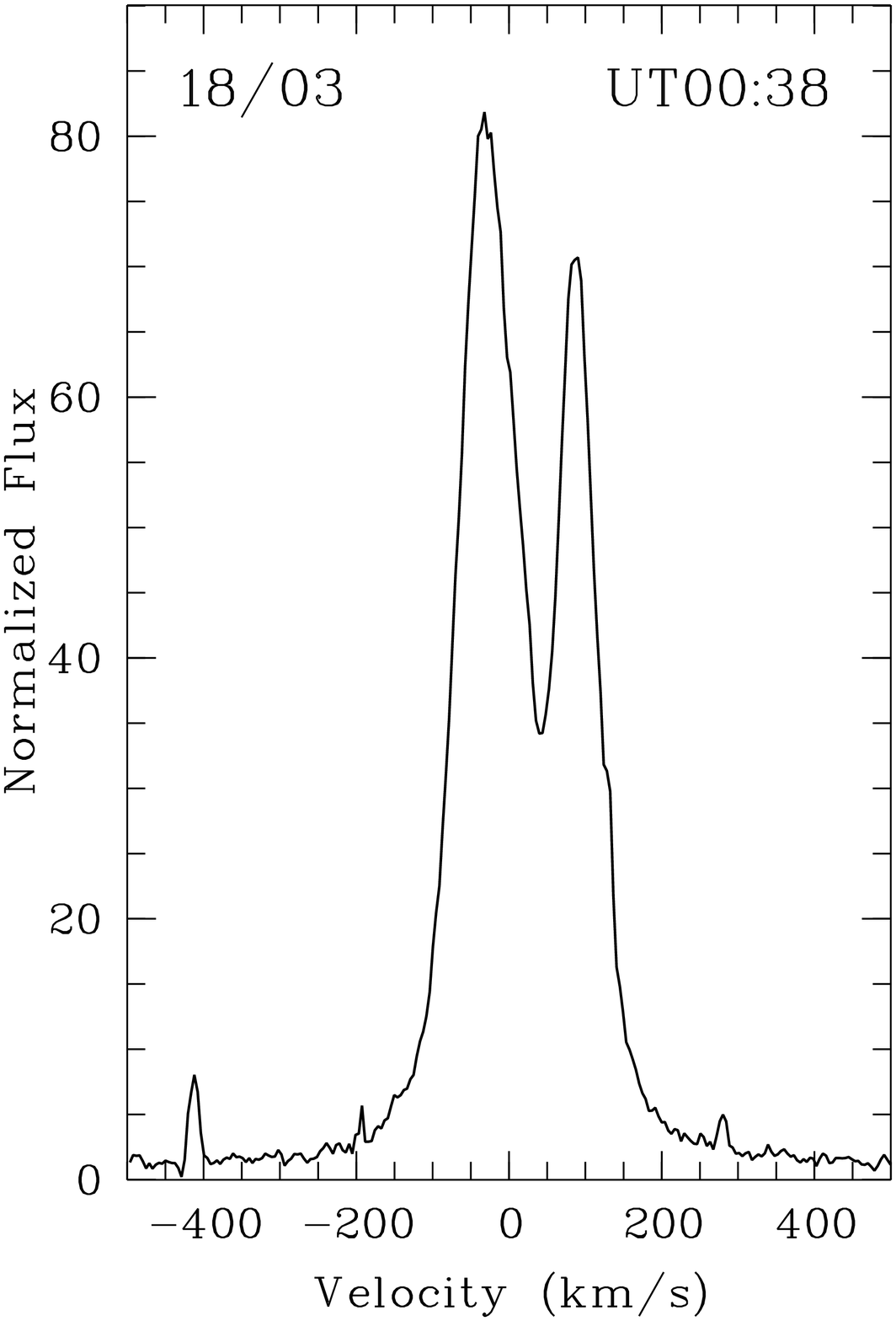}
\includegraphics[width=3.1cm,height=3.5cm]{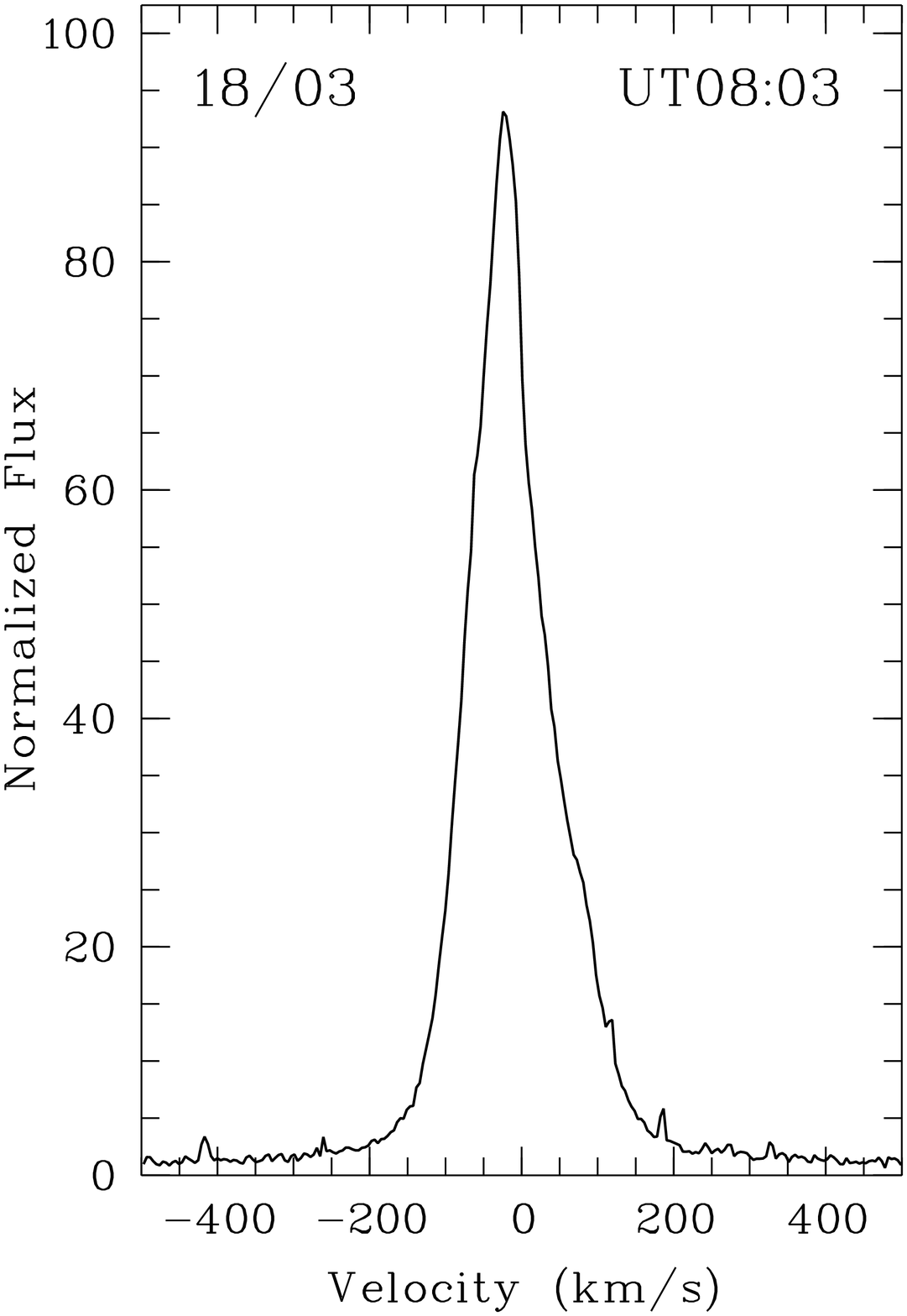}
\includegraphics[width=3.1cm,height=3.5cm]{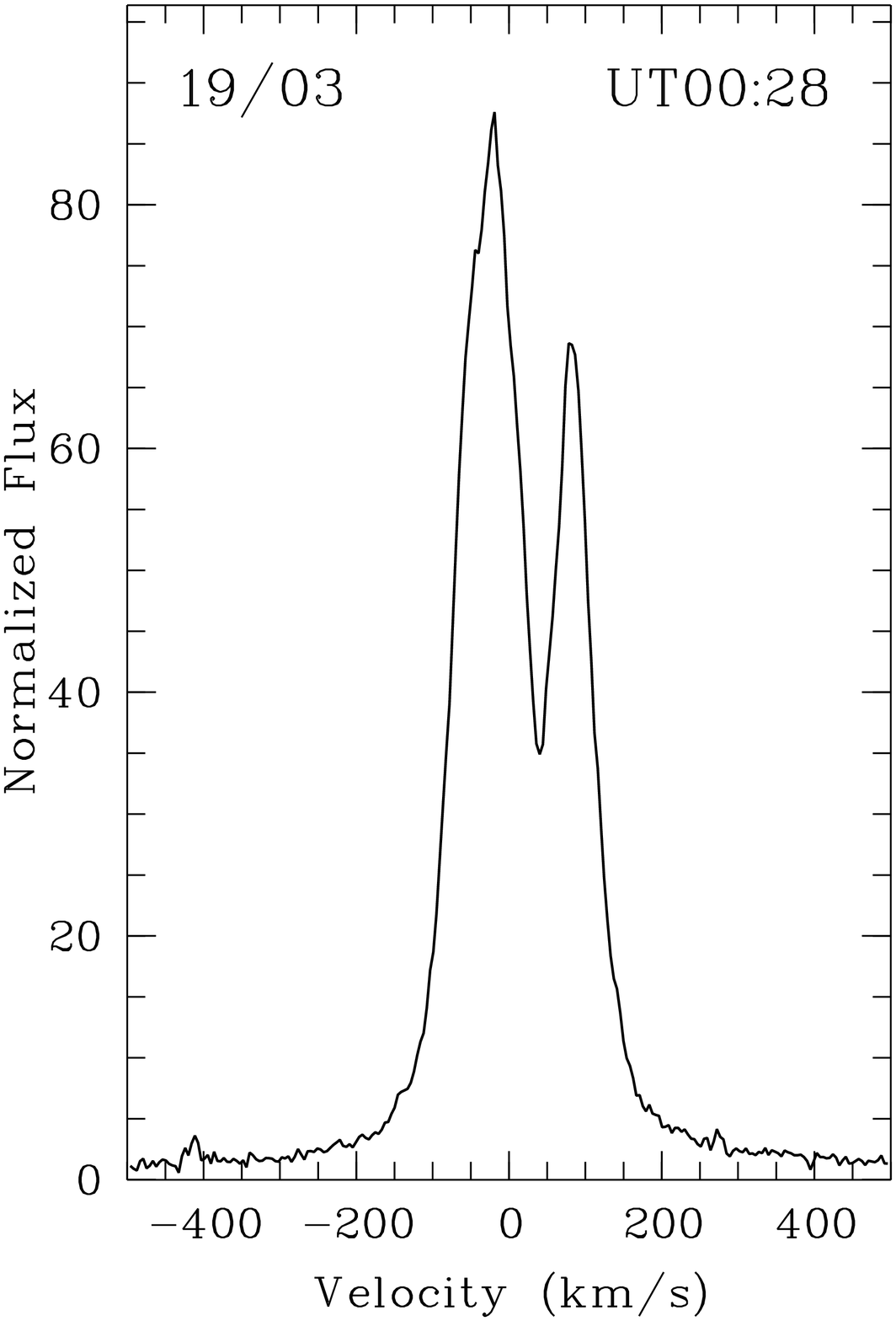}
\includegraphics[width=3.1cm,height=3.5cm]{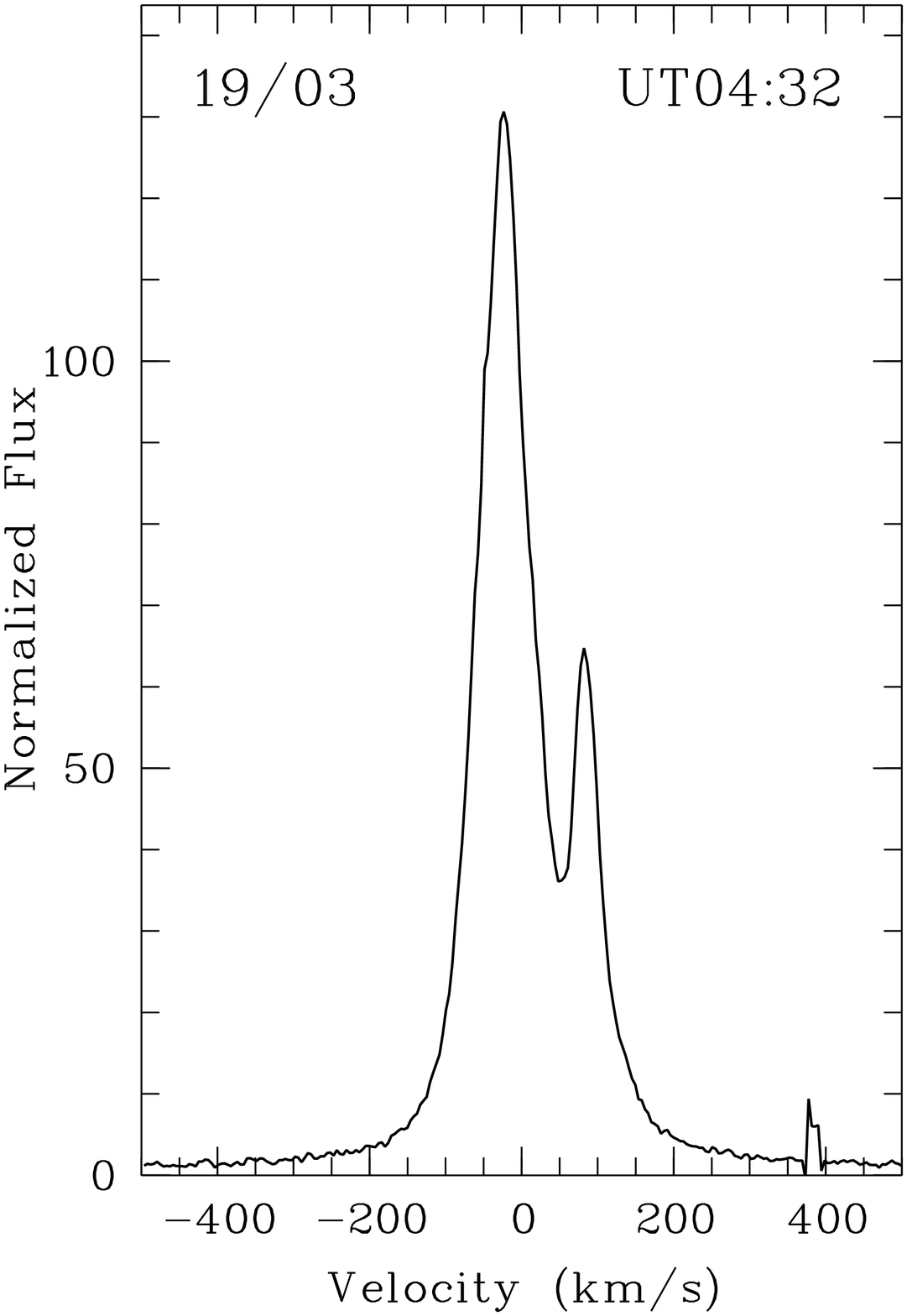}
\includegraphics[width=3.1cm,height=3.5cm]{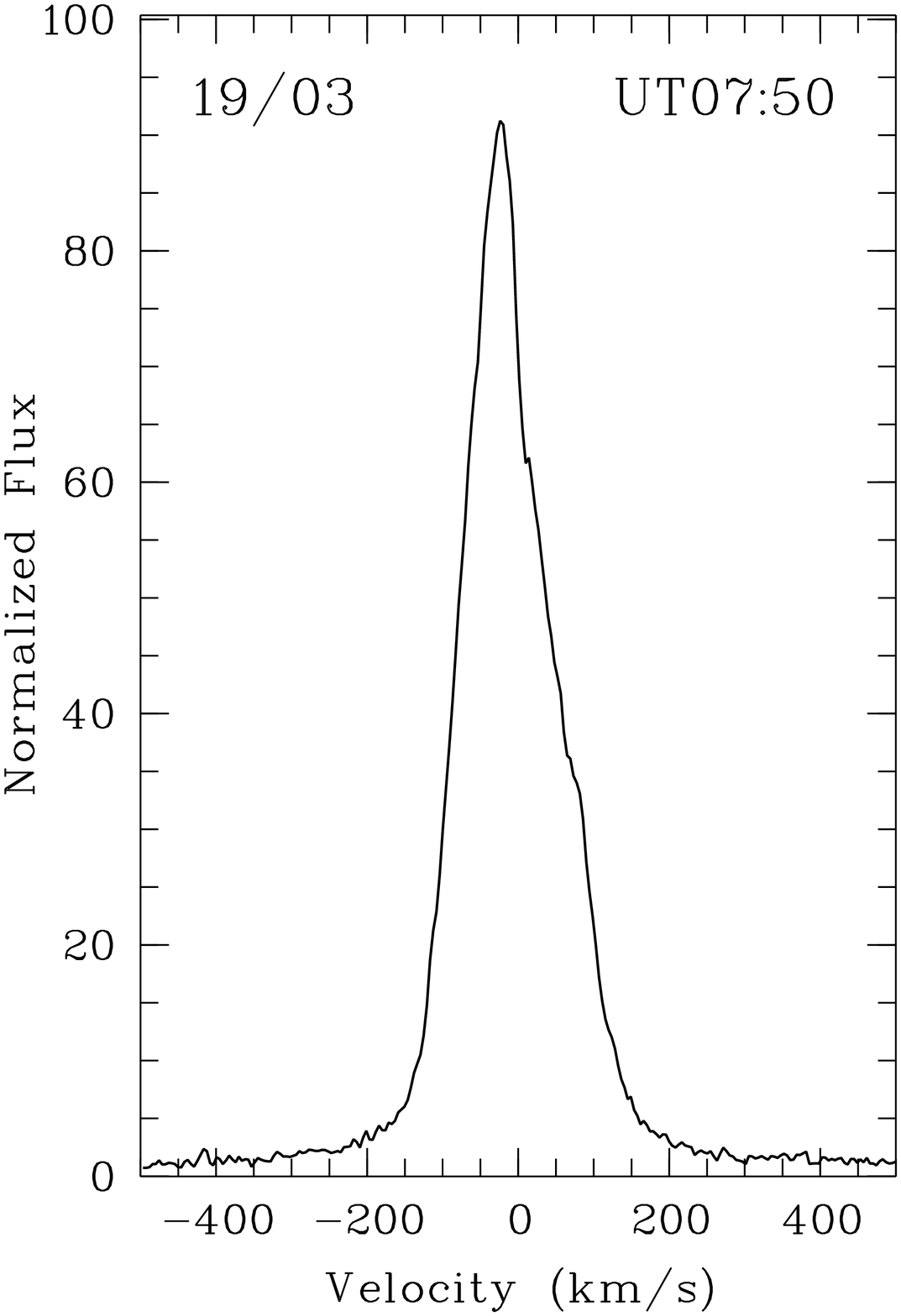}\\
\includegraphics[width=3.1cm,height=3.5cm]{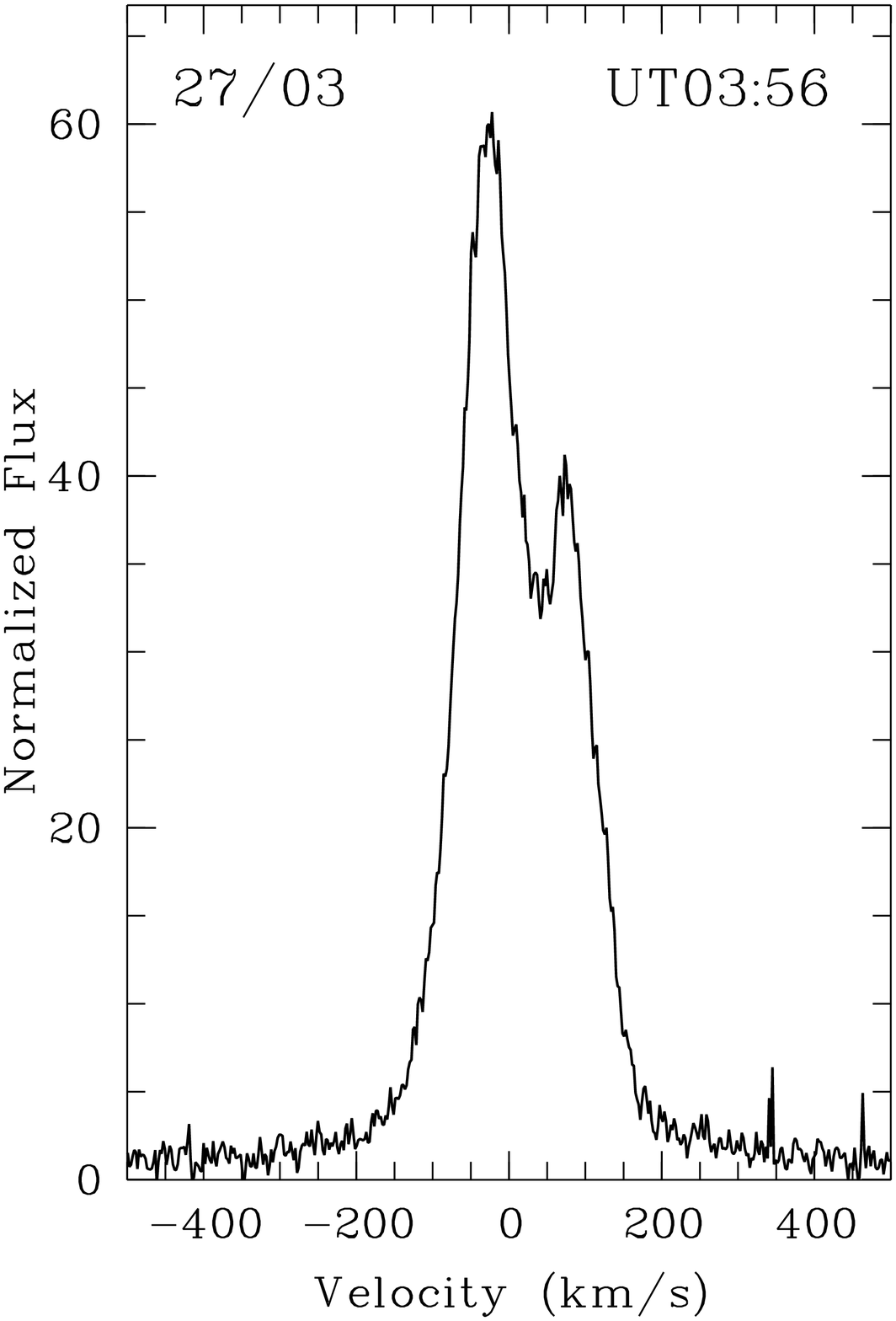}
\includegraphics[width=3.1cm,height=3.5cm]{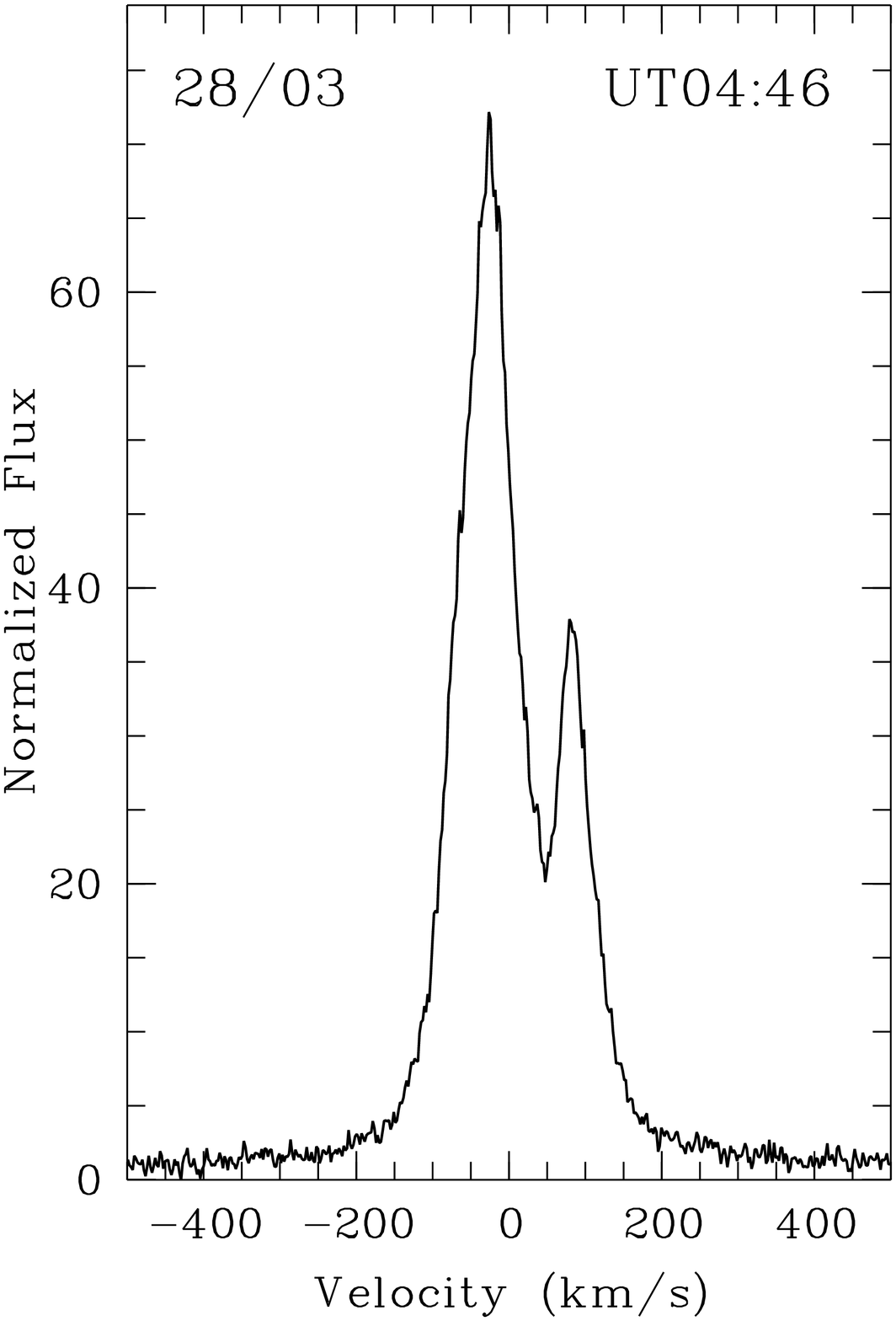}
\includegraphics[width=3.1cm,height=3.5cm]{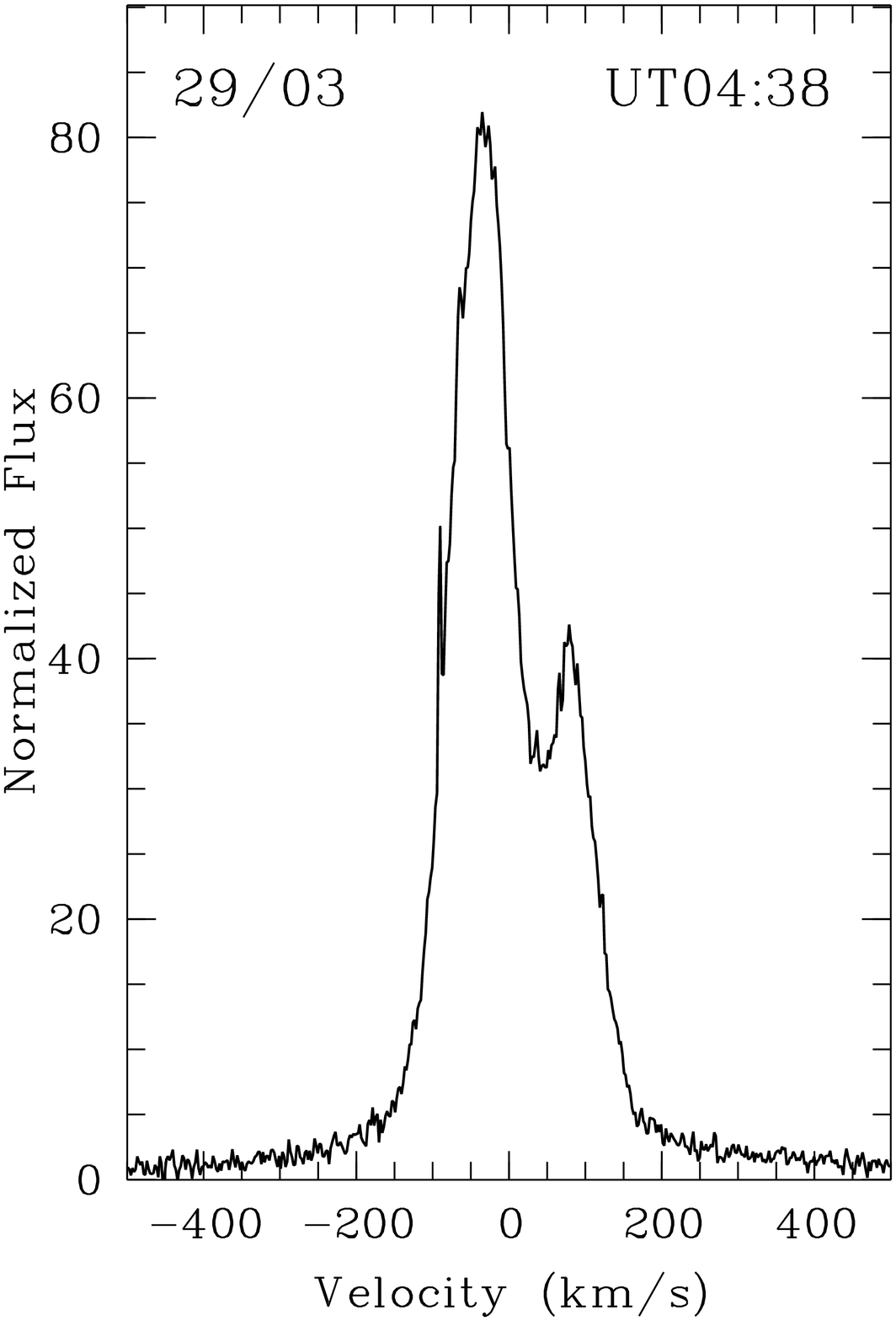}
\includegraphics[width=3.1cm,height=3.5cm]{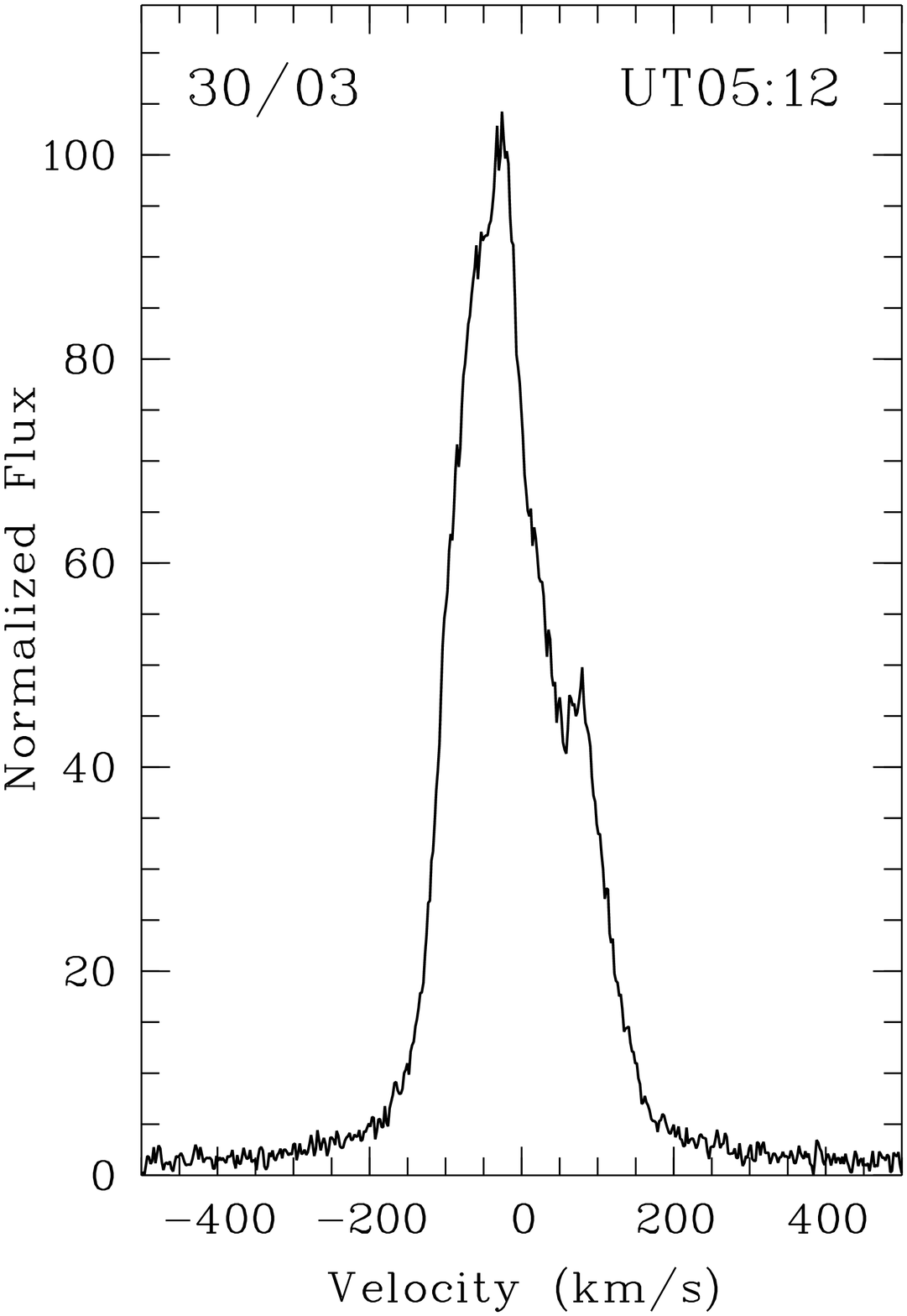}
\includegraphics[width=3.1cm,height=3.5cm]{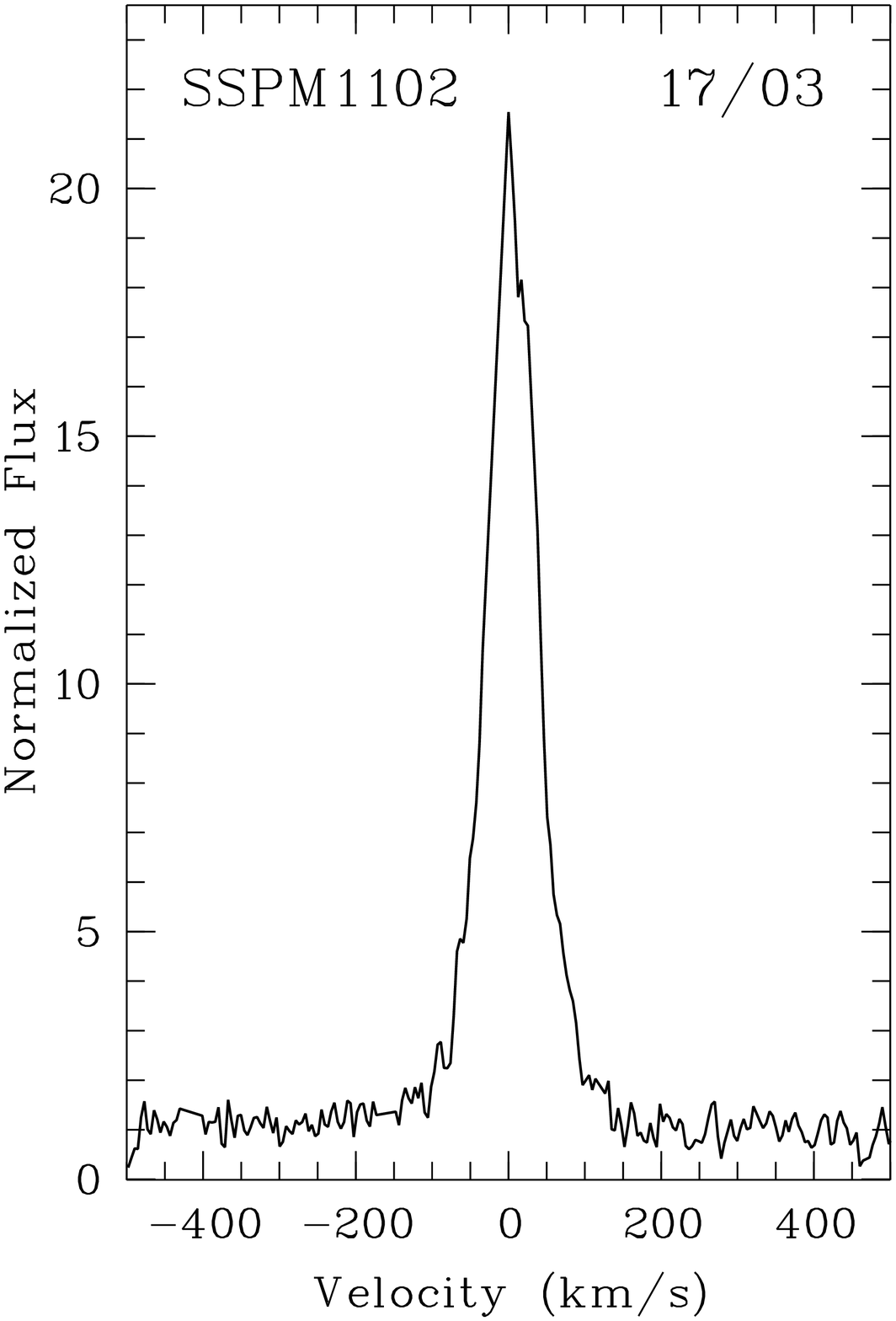}
\caption{H$\alpha$ line profiles for 2M1207 (panels 1-14) and SSPM1102 (panel 15). 
The spectra of 2M1207 are ordered according to Table \ref{log}. All fluxes are 
normalized to the continuum. \label{halpha}}
\end{center}
\end{figure}

The spectra of 2M1207 show a variety of emission lines. The most prominent features 
are the Hydrogen Balmer lines, from H$\alpha$ down to H10 at 3798\,\AA. HeI emission 
lines are clearly seen at 5876, 6678, and 7065\,\AA. In 
addition, we detect emission in NaD (5890+5895\,\AA), OI (7774 and 8446\,\AA), several 
H Paschen lines, and in CaH+K (3968+3933\,\AA). The distinctive emission line spectrum 
appears in all epochs, but the lines are considerably weaker in the January data compared 
to the March run. In the March spectra, the Balmer and HeI lines exhibit broad wings 
and are in most cases clearly asymmetric. In contrast, in the January data these lines 
are much narrower and more symmetric. We note that none of our spectra shows forbidden 
line emission in [OI] or [SII]. The LiI 6708\,\AA\ absorption 
is detected in the co-added 2M1207 spectrum with an EW of 
$0.4\pm0.1$\,\AA, in agreement with the value of 0.5 given by MJB03.  
Our second target SSSPM1102 exhibits 
similar emission features to 2M1207, in particular H Balmer and HeI, but the lines 
are narrow and fairly symmetric (Fig. \ref{halpha}, last panel). In that sense, 
SSPM1102 shows a spectrum comparable to 2M1207 during the January run. Lithium
is tentatively detected in SSPM1102, with an EW of $0.4\pm 0.2$\,\AA, consistent 
with membership in the TW Hydrae group.

In this {\it Letter}, we focus on the most prominent H Balmer features of 2M1207, 
particularly H$\alpha$, which shows dramatic changes both in shape and intensity 
in our dataset. In Fig. \ref{halpha} we present the complete time series of 
H$\alpha$ profiles. In the three January spectra (first three panels), the line is 
relatively weak and only slightly asymmetric. The March spectra, however, show very 
intense and in most cases strongly asymmetric H$\alpha$ features. Furthermore, the 
profile exhibits dramatic, quasi-periodic changes, from strongly double-peaked to 
nearly symmetric with only a weak red 'shoulder', over the course of a night. 
The profile shapes of H$\alpha$ in the March spectra can be understood  
as a superposition of a broad emission peak and a shallower absorption feature. The
absorption component is clearly redshifted, with an offset of 40-60\,$\mathrm{km\,s^{-1}}$ 
(see Table \ref{log}). Absorption and emission are both variable; the 
absorption component disappears and re-appears on timescales of about 1\,day. 

For the H$\alpha$ emission line, which is the dominant feature in the red part of 
the spectrum, we determined the full width at 10\% of the peak (hereafter 10\% width,
see Table \ref{log}). In addition, we measured line fluxes for prominent features by 
integrating the spectrum after subtracting the continuum. In the absence of absolute 
flux calibrations, these line fluxes can only be compared to values measured in the 
same spectrum. In Table \ref{log} we list the flux ratio between the H$\beta$ and 
the H$\alpha$ Balmer lines. The results are given in Table \ref{log}. The same 
measurements were also carried out for SSSPM1102.

\section{Geometry of the system}

Many of the emission lines seen in the spectrum of 2M1207 are typical indicators
for on-going accretion, and are frequently observed in classical T~Tauri stars (CTTS). 
Perhaps the most intriguing feature is the redshifted absorption component in the 
H$\alpha$ profile, which shows clear changes within
a few hours. Redshifted absorption is often seen in the emission lines 
of CTTS, such as AA Tau, BP Tau \citep{jb95}, DN Tau \citep{mhc98}, and TW Hya 
\citep{ab02}. These features are clear signatures of infalling material, and 
are usually explained in a scenario in which the line of sight points into a funnel 
flow, and we see the hot spot(s) against the inflowing gas and not against the 
photosphere. According to the magnetospheric accretion models of \citet{hhc94}, 
red absorption is a strong function of the inclination $i$ between the rotational 
axis and the line of sight, and should only appear if the system has 
$i\gtrsim 60^{\circ}$, i.e. when the disk is seen close to edge-on. Indeed, 
it has been shown that T~Tauri stars with red absorption have disks at fairly high 
inclinations, e.g. AA Tau \citep[$i=75^{\circ}$,][]{bca99} and BP Tau 
\citep[$i=60-70^{\circ}$,][]{sdg00}. Therefore, it is likely that 2M1207 is also 
seen with an inclination $i\gtrsim 60^{\circ}$, i.e. close to edge-on.

In this scenario, one would expect the red absorption to vary on timescales similar 
to the rotation period, as a consequence of hot spots co-rotating with the object. 
MJB03 determined a projected rotational velocity ({\it v~sin~i}) of 
13$\,\mathrm{km\,s^{-1}}$ for 2M1207, which corresponds to an upper period limit 
of 25 h, assuming a radius of $\sim 0.27\,R_{\odot}$ for 0.03$\,M_{\odot}$ 
and 10\,Myr \citep{cba00}.\footnote{The radius, and thus the upper period 
limit, keeps roughly constant for reasonable uncertainties in mass (30\%) and 
age (5-20\,Myr).} If the system is seen at high inclination, the 
rotation period is in the range of one day. Thus, the red absorption feature in 2M1207 
is indeed variable on timescales of the rotation period. Since this is difficult to 
explain for a system seen close to face-on, this provides further evidence for our given 
interpretation. The reported lack of $K-L$ near-infrared colour excess in 2M1207 
\citep{jas03} may also be consistent with a disk at high inclination, though alternate 
explanations, such as grain growth or an inner disk hole, are also plausible. 

As pointed out in Sec. \ref{obs}, 2M1207 shows the complete Balmer series
down to 3800\,\AA. In spectra with a nearly symmetric H$\alpha$ profile, 
the line fluxes in the higher Balmer lines tend to be increased relative to 
H$\alpha$ (Table \ref{log}). This tendency provides additional 
evidence for our given interpretation of the profile changes: If H$\alpha$ is 
double-peaked, the absorption dominates, and we see on average cooler gas than in 
the symmetric case, where the hot spot is unobscured.

We conclude that the most straightforward explanation for the shape and
variations of the H$\alpha$ profiles is a high inclination, implying that
we see the disk close to edge-on. This explains both the presence of strong 
redshifted absorption as well as its variation on timescales of the rotation
period. At times when we see strong redshifted absorption, we may be 
looking into an accretion column. The fact that we probably see rotational
modulation of this absorption provides evidence for inhomogenuous rather
than spherical accretion, as expected in a scenario where the accretion
is channeled by the magnetic field. Thus, our findings strongly support 
magnetospheric accretion in brown dwarfs.

\section{Variable Accretion in 2M1207}

\begin{figure}[t]
\begin{center}
\includegraphics[width=5.5cm,angle=-90]{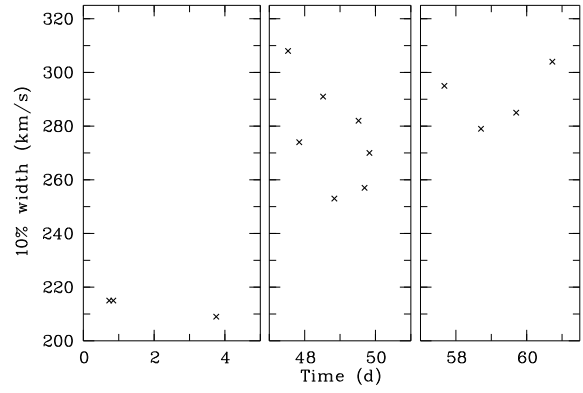}
\caption{10\% width of H$\alpha$ as a function of time (Julian Date 
minus 2453399.0). The three panels show the data from three different
observing runs. A significant increase of the H$\alpha$ width was observed
between the first and the second run.\label{ten}}
\end{center}
\end{figure}

By comparing the spectra from January with those from March, it is obvious
that the mass accretion process on 2M1207 is significantly variable. There
is a distinct change in the line profiles (see Fig. \ref{halpha}) and 
an increase in the 10\% width, which is shown as a function of time in 
Fig. \ref{ten}. In the March spectra, the H$\alpha$ profile is clearly 
dominated by accretion, with broad emission and additional redshifted 
absorption. The higher levels of the Balmer series and HeI are also 
broad and asymmetric, again indicating on-going accretion 
\citep{mhc03}. The 10\% widths in March are substantially higher than 
200$\,\mathrm{km\,s^{-1}}$ (Fig. \ref{ten}), which has been proposed 
as a useful empirical threshold between accreting and 
non-accreting systems \citep{jmb03}. In contrast, the three
January spectra are significantly narrower, and only slightly asymmetric.
In one January spectrum, weak blue-shifted absorption is visible, usually an 
indication of outflowing material. Given the lack of forbidden line emission, 
this is the only evidence for wind in our spectra. The 10\% width in January 
is only slightly larger than 200$\,\mathrm{km\,s^{-1}}$, which implies that 
2M1207 is only weakly accreting, if at all, at that time. 

The 10\% width of H$\alpha$ can be used to estimate the accretion rate. 
\citet{ntm04} calibrated the correlation between accretion rate and 10\% width 
using objects for which the accretion rate was derived from continuum veiling. 
We now use their empirical relation to assess the accretion rate for 2M1207. 
For the three January spectra, we obtain an average value 
of $213\pm 3\,\mathrm{km\,s^{-1}}$ for the 10\% width. This corresponds to 
an accretion rate of $10^{-10.8\pm 0.5}\,M_{\odot} \mathrm{yr}^{-1}$, where the error 
estimate takes into account both the scatter of the 10\% widths and the uncertainty in  
transforming to accretion rates. In contrast, the spectra from March 2005 
show a 10\% width of $281\pm 18\,\mathrm{km\,s^{-1}}$. This converts to an accretion 
rate of $10^{-10.1\pm 0.7}\,M_{\odot} \mathrm{yr}^{-1}$. Although the accretion rates 
in January and March are comparable within the errors\footnote{We note, however, that the 
uncertainty in the relation between accretion rate and 10\% width is mainly caused 
by differences in individual objects, whereas we consider here only one particular 
object. Thus the difference between January and March accretion rates is 
more significant than the formal error suggests.}, the 10\% widths are higher in March 
by 32\%, which clearly points to a significant change in the accretion rate as 
well. 

It is instructive to compare these results with previous spectroscopic
observations of 2M1207. \citet{g02} report very strong H$\alpha$ emission 
in two consecutive spectra for 2M1207 taken in 2001, with EWs 
$\sim 300$\,\AA, corresponding to 10\% widths in the range of 300$\,\mathrm{km\,s^{-1}}$ 
(Muzerolle et al. 2003), i.e. comparable to our March data. On the other hand, 
MJB03 and Kirkpatrick et al. \citep[unpublished, see][]{gb04} have measured 
H$\alpha$ EWs of 28 and 42\,\AA, likely corresponding to 10\% widths 
around 200$\,\mathrm{km\,s^{-1}}$, i.e. similar to our January spectra. Thus, the 
literature data support the strong variations in the H$\alpha$ widths (and hence
the accretion rate). From the available data we conclude that the 
accretion rate of 2M1207 changes by a factor of 5-10 on a timescale
of six weeks. Strong variations in the accretion behaviour have been observed for 
T~Tauri stars in the past. \citet{bb90} find veiling variations 
in the range of an order of magnitude for several objects, e.g. RW Aur, AA Tau,
DG Tau, which implies significant changes in the accretion rate \citep{mhc98}. 
The veiling of TW Hya is variable between 0.84 
and 0.12 on timescales of a few months \citep{ab02}, again suggesting 
variability in the accretion rate. Here we confirm for the first 
time that brown dwarfs can behave in a very similar manner. Observations with 
better time sampling and longer baselines are required to obtain more precise 
constraints on the timescales and the extent of the accretion variability of 2M1207.

As already mentioned, our second target SSSPM1102 shows similar
spectral features to 2M1207, in particular H Balmer lines. Its H$\alpha$ 
10\% width is 194$\,\mathrm{km\,s^{-1}}$, and the profile shows a weak
red absorption component (see Fig. \ref{halpha}, last panel). Thus, according 
to the criteria of \citet{jmb03}, SSSPM1102 is only weakly accreting, if at all,
at the time of observation. The H$\alpha$ profile shape and 10\% width are 
similar to the January data of 2M1207, indicating an accretion rate of 
$\sim 10^{-11}\,M_{\odot}\mathrm{yr}^{-1}$ for SSSPM1102.

\section{Conclusions}

We have obtained a time series of 14 high-resolution spectra of the young, accreting
brown dwarf 2M1207, which show a variety of emission lines confirming on-going 
accretion. These features are, however, significantly weaker in the January 
spectra in comparison to the March data. Particularly the H$\alpha$ line is
strongly variable in shape and width. The 10\% width varies by 32\% on a timescale
of six weeks, indicating a change in the accretion rate by a factor of 5-10. 
In the high accretion state, the line profile changes from double-peak to nearly 
symmetric on timescales of a few hours, with a redshifted absorption feature that 
disappears and re-appears on timescales of $\sim 1$\,d, consistent with the brown 
dwarf rotation period. This profile variability can be explained in a scenario where 
the accretion disk is seen at a relatively low inclination with respect to the 
line of sight. In cases where we see a double-peaked H$\alpha$ line, we may be 
looking into an accretion column, co-rotating with the object. Thus, the 
profile variability provides strong evidence for funneled rather than spherical 
accretion. Hence, these observations make a compelling case that large-scale 
magnetic fields play an important role in the accretion process of brown dwarfs.

\acknowledgments
We thank the Las Campanas Observatory staff for their assistance and the anonymous
referee for helpful suggestions. This research was supported by an NSERC grant and 
University of Toronto startup funds to R.J.

\clearpage

\clearpage

\clearpage

\begin{table}
\begin{center}
\begin{tabular}{cccccc}
\tableline\tableline
\noalign{\smallskip}
No. & Date & UT & H$\alpha$10\% & H$\beta$/H$\alpha$ & $\Delta V_\mathrm{Abs}$\\
    & 2005 &    & ($\mathrm{km\,s^{-1}}$) & & ($\mathrm{km\,s^{-1}}$) \\          
\noalign{\smallskip}
\tableline
\noalign{\smallskip}
1 & 29/01  & 05:25 & 215 & 0.30 & \\
2 & 29/01  & 08:08 & 215 & 0.42 & \\
3 & 01/02  & 05:56 & 209 & 0.38 & \\
4 & 17/03  & 00:38 & 308 & 0.21 &  46 \\
5 & 17/03  & 08:14 & 274 & 0.31 & \\
6 & 18/03  & 00:23 & 291 & 0.14 &  40 \\
7 & 18/03  & 08:03 & 253 & 0.30 & \\
8 & 19/03  & 00:28 & 283 & 0.38 &  40 \\
9 & 19/03  & 04:32 & 257 & 0.35 &  49 \\
10 & 19/03 & 07:50 & 270 & 0.36 & \\
11 & 27/03 & 03:56 & 295 & 0.23 &  41 \\
12 & 28/03 & 04:46 & 279 & 0.22 &  47 \\
13 & 29/03 & 04:38 & 285 & 0.25 &  41 \\
14 & 30/03 & 05:12 & 304 & 0.27 &  59 \\
\noalign{\smallskip}
\tableline\tableline
\noalign{\smallskip}
\end{tabular}
\caption{Observing log and line data for 2M1207. The last column contains the velocity
offset of the red absorption component, if it is clearly detectable. The error for 
velocities is $\sim 10\,\mathrm{km\,s^{-1}}$, for flux ratios $\sim 10\%$. \label{log}}
\end{center}
\end{table}

\end{document}